\documentstyle[12pt]{article}         
\normalbaselines                      
\begin{document}                      
\bibliographystyle{unsrt}             
\def\beee{\begin{equation}}           
\def\eeee{\end{equation}}             
\def\dggg{^{\dagger}}                 
\vbox{\vspace{6mm}}                   
\begin{center}
{\large \bf Spin-Statistics, Spin-Locality, and TCP:\\
Three Distinct Theorems}\\[.1in]

O.W. Greenberg\footnote{Supported in part by a Semester Research Grant from the 
General Research Board of the University of Maryland and by a
grant from the National Science Foundation.}\\
Center for Theoretical Physics\\
Department of Physics\\
University of Maryland\\
College Park, Maryland 20742-4111\\[.4in]
University of Maryland Preprint No. 98-002\\
hep-th/9707220
\end{center}

\begin{abstract}
I show that the spin-statistics theorem has been confused with another theorem
that I call the spin-locality theorem.  I also argue that the spin-statistics 
theorem properly depends on the properties of asymptotic fields which are free 
fields.  In addition, I discuss how ghosts evade both theorems, give the basis
of the spin-statistics theorem for fields without asymptotic limits such as 
quark and gluon fields, and emphasise the weakness of the requirements for the 
$TCP$ theorem.
\end{abstract}

\vspace{.25in}

I have two purposes in this note.  The first is to make clear the difference 
between the spin-statistics theorem: {\it particles} that obey Bose
statistics must have integer spin and {\it particles} that obey Fermi
statistics must have odd half-integer spin\cite{mf,wp}, and what I suggest 
should be 
called the spin-locality theorem: {\it fields} that commute at spacelike 
separation must have integer spin and
{\it fields} that anticommute at spacelike separation must have odd
half-integer spin[3-7].    
My second purpose is to emphasize the weakness of the conditions
under which the $TCP$ theorem\cite{jost} 
holds and in that way to distinguish it from the
spin-statistics theorem and the spin-locality theorem.  In doing this I 
amplify Res Jost's example\cite{jost2} of a
field that has the wrong spin-statistics connection, but obeys the $TCP$
theorem.  The thrust of this note is to separate these
three theorems that are sometimes lumped together.

{\flushleft 1. Spin-statistics and spin-locality}

Since the ``right'' cases of both the spin-statistics theorem and the
spin-locality theorem agree, I emphasize what fails in each of the theorems
for the ``wrong'' cases.  Spacelike commutativity (locality) of observables 
fails
for the wrong cases of the spin-statistics theorem.  For example, as I will
discuss in detail in amplifying Jost's example, for 
a neutral spin-0 scalar field
that obeys Fermi statistics, observables, such as currents, fail to be
local.  By contrast, a neutral spin-0 scalar field whose anticommutator is local
does not exist--it is identically zero.  The obvious corresponding wrong cases
for spin-1/2 and higher spin have the corresponding failures.\\

{\it  Spin-statistics:}~~
Because the spin-statistics theorem refers to the statistics of particles, its
formulation in field theory should involve the operators that create and
annihilate particles.  These operators are the asymptotic fields, the in- and 
out-fields.  
Since the asymptotic fields (at least for massive particles) are free
fields, the proof of the spin-statistics theorem only requires using the
properties of free fields.  The assumptions necessary for the proof are (1)
that the
space of states is a Hilbert space, i.e., the metric is positive-definite,
(2) the fields smeared with test functions in the Schwartz space ${\cal S}$ have
a common dense domain in the Hilbert space, (3) the fields transform under a
unitary representation of the restricted inhomogeneous Lorentz group, (4) the
spectrum of states contains a unique vacuum and all other states have positive
energy and positive mass, and (5) the bilinear observables constructed from the
(free) asymptotic fields commute at spacelike separation (local commutativity
of observables).  Using these assumptions for free fields of any spin, 
Fierz\cite{mf} and 
Pauli\cite{wp} proved that integer-spin particles must be bosons and odd
half-integer spin particles must be fermions.  They used locality of observables
as the crucial condition for integer-spin particles 
and positivity of the energy as the crucial
condition for the odd half-integer case.  Weinberg\cite{wein} 
showed that one can use the locality of
observables for both cases if one requires positive-frequency modes to be
associated with annihilation operators and negative-frequency modes to be
associated with creation operators.  \\

I assume 
that for non-gauge theories with no massless particles the asymptotic fields are
an irreducible set of operators.  I show that the conserved
observables such as the energy-momentum operators and, for theories with
conserved currents, the current operators, must be a sum of the free field
functionals of the asymptotic fields, where the sum runs over the independent
asymptotic fields, including those for bound states if there are bound states in
the theory.  To see this, require--say for the in-fields--
\beee
i[P^{\mu}, \phi^{in}(x)]_-=\partial^{\mu} \phi^{in}(x)             \label{1}
\eeee
for the case of the energy-momentum operator and a neutral scalar field.  The
general expansion in the in-fields for $P^{\mu}$ is
\beee
P^{\mu}=\sum_{n=0}^{\infty}\frac{1}{n!}\int f^{(n)~\mu}(x_1, \cdots, x_n)
\stackrel{\leftrightarrow}{\partial}^{\mu_1}_{x_1} \cdots
\stackrel{\leftrightarrow}{\partial}^{\mu_n}_{x_n}
:\phi^{in}(x_1) \cdots \phi^{in}
(x_n):d\Sigma_{\mu_1}(x_1) \cdots d\Sigma_{\mu_n}(x_n)            \label{1.1}
\eeee
and inserting it in 
Eq.(\ref{1})
shows that only the constant term and the bilinear free functional of the 
in-fields can enter $P^{\mu}$. The requirement that the vacuum have zero energy
eliminates the constant term.  The equation for the bilinear term is
\beee
\int f^{(2)}(x_1,x)\stackrel{\leftrightarrow}{\partial}^{\mu_1}_{x_1}:
\phi^{in}(x_1):d\Sigma_{\mu_1}(x_1)=\partial^{\mu}\phi^{in}(x).     \label{1.2}
\eeee
The integrals over the spacelike surfaces $\Sigma(x_i)$ are independent of the
time because of the time-translation invariance of the Klein-Gordon scalar
product.  Thus the solution of Eq.(\ref{1.2}), 
\beee
f^{(2)}(x_1,x)=-\partial^{\mu}_x \Delta(x-x_1),                  \label{1.3}
\eeee
leads to the usual result for $P^{\mu}$ using $\Delta(0,{\bf x})=0$, 
$\partial_0\Delta(0,{\bf x})=-\delta({\bf x})$, 
and the Klein-Gordon equation for $\Delta(x)$.
Thus the arguments of Fierz, Pauli, and Weinberg for free fields hold in 
the case of interacting theories
that have an irreducible set of in- (or out-) fields.
For example the charge density for a charged spin-zero field is 
\beee
j^{\mu}(x)=i:
\phi^{as ~\dagger}(x)\stackrel{\leftrightarrow}{\partial}^{\mu}\phi^{as}(x):
                                                             \label{2}
\eeee
and for
a charged spin-one-half field it is 
\beee
j^{\mu}(x)=:{\bar \psi}^{as}(x)\gamma^{\mu}
\psi^{as}(x):.                                               \label{3}
\eeee
  For a spin-zero field, the commutator of the currents 
$[j^{\mu}(x),j^{\nu}(y)]_{-}$ will
contain the local distribution $i\Delta(x-y)$ if the annihilation and creation
operators obey Bose commutation relations and the nonlocal distribution
$\Delta^{(1)}(x-y)$ if the annihilation and creation operators obey Fermi
commutation relations.  For a spin-one-half field, the commutator $[j^{\mu}(x),
j^{\nu}(y)]_{-}$ will contain the local distribution $iS(x-y)$ if the particle
operators obey Fermi rules and the nonlocal distribution $S^{(1)}(x-y)$ if the 
particles obey Bose rules.  (Here I assume that the fields are expanded in
annihilation operators for the positive frequency modes and in creation
operators for the negative frequency modes.  If a Dirac field is expanded in
annihilation operators for both types of modes, then the commutator of the field
and its Pauli adjoint will be the local distribution $iS(x-y)$, but the energy
operator will be unbounded below.\cite{pauli})\\

{\it Spin-locality:}~~
For the spin-locality theorem, L\"uders and Zumino\cite{lz} and
Burgoyne\cite{bur} replaced assumption (5) of the spin-statistics theorem by
(5$^{\prime}$) that the fields either commute
\beee
[A_{\mu}(x),A^{\dagger}_{\nu}(y)]_-=0, (x-y)^2<0,       \label{4}
\eeee   
or anticommute
\beee
[\psi_{\alpha}(x),\bar{\psi}_{\beta}(y)]_+=0, (x-y)^2<0.      \label{5}
\eeee
at spacelike separation. Here $[A,B]_{\pm}=AB \pm BA$.
 Since in general the fields are not observables, 
this is not an assumption about physical quantities.
I will call such fields local or antilocal and, as mentioned above, I will
call the theorem the spin-locality theorem.   
The L\"uders-Zumino and Burgoyne proof shows that if the fields have the wrong 
commutation
relations, i.e. integer-spin fields are antilocal and
odd-half-integer-spin fields are local, the fields vanish.    
This assumption does not relate 
directly to particle statistics and for that reason this theorem should not 
be called the spin-statistics theorem.  Thus the assumptions of the 
spin-statistics theorem and of the spin-locality theorem
differ; further, the conclusions of the two theorems differ for the case of 
the wrong association between spin and either statistics or type of locality.\\

{\it Ghosts:}~~
There is a case in practical calculations in which both 
the spin-statistics theorem and the spin-locality theorem seem to be
violated: namely, the ghosts of gauge theory.  These are scalar fields that (1)
anticommute at spacelike separation and thus seem to violate the spin-locality
theorem and (2) whose
asymptotic limits are quantized obeying Fermi particle statistics and seem to 
violate the spin-statistics theorem.  Most discussions of gauge theory rely on
path integrals and don't explicitly consider the commutation or anticommutation
relations of ghost fields.  N. Nakanishi and I. Ojima\cite{no} give
\beee
[C^{as}_i(x),{\bar C}^{as}_j(y)]_+=-\delta_{ij}D(x-y),
\eeee
where $i$ and $j$ run 
over the adjoint representation of the gauge group,
as the anticommutator between the ghost and antighost fields.  The
anticommutators of $C^{as}$ with itself and of ${\bar C}^{as}$ with itself
vanish.  The arguments used in the proof of the spin-locality theorem show that
the two-point functions $\langle 0|C(x) C(y)|0\rangle$ and
$\langle 0|{\bar C}(x){\bar C}(y)|0\rangle$ both vanish.  Since $C$ and ${\bar
C}$ are hermitian, if the metric of the space were positive-definite the fields
$C$ and ${\bar C}$ would annihilate the vacuum and the fields would vanish.
Because the space of states is indefinite, this conclusion does not follow.
The off-diagonal form of these anticommutators is connected with the
fact that the ghost and antighost fields create zero norm states.  Nakanishi and
Ojima take the ghost and antighost fields to be independent hermitian fields, so
the assumptions of neither the spin-statistics nor of the spin-locality theorem
hold and there is no violation of either theorem.\cite{no}  I give the 
anticommutation relations of Nakanishi and Ojima for the asymptotic fields for 
the annihilation and 
creation operators of the ghosts and antighosts to illustrate from the particle
point of view how these fields evade the two theorems,
\beee
[C^{(as)}(k),{\bar C}^{(as)~\dagger}(l)]_+=i 2E_k \delta({\bf k}-{\bf l}), ~~
[{\bar C}^{(as)}(k),C^{(as)~\dagger}(l)]_+=-i 2E_k \delta({\bf k}-{\bf l}),
\eeee
where other anticommutators vanish, and 
I used relativistic normalization for the annihilation and creation
operators,
\beee
C^{(as)}(x)=\frac{1}{(2 \pi)^{3/2}}\int \frac{d^3k}{2E_k}[C^{(as)}(k)
exp(-i k \cdot x)+C^{(as)~\dagger}(k)exp(i k \cdot x)]
\eeee
and a similar formula for ${\bar C^{(as)}}$.  
These two anticommutators go into each
other under hermitian conjugation.  The $i$ factors are what allow the
anticommutator $[C^{(as)}(x),{\bar C^{(as)}}(y)]_+$ to be $-D(x-y)$ 
(for the massless case),
rather than a multiple of $D^{(1)}(x-y)$.\\

{\it Fields without an asymptotic limit:}~~
For fields that do not have asymptotic fields, such as quark or gluon fields, 
one needs a condition on the fields that can replace the condition on the
asymptotic fields in deriving the spin-statistics theorem.  I have
argued\cite{thesis} that the c-number equal-time canonical commutation 
(anticommutation)
rules for the fields lead to the commutation (anticommutation) relations for the
asymptotic fields using the LSZ weak asymptotic limit.  This suggests that the
requirement of local commutativity of observables that is
satisfied by asymptotic fields by having either Bose or Fermi statistics can be
satisfied for fields that do not have asymptotic fields by having either the 
canonical equal-time commutators or the canonical equal-time 
anticommutators to be c-numbers.  This alternative replaces the alternative of 
either locality or antilocality of the fields 
of the L\"uders-Zumino and Burgoyne theorem.  
The commutator of currents
at equal times will involve a sum of terms with either equal-time commutators or
equal-time anticommutators of the fields.  Since these are c-numbers, they 
will vanish,
except at coincident points, only if they have the correct choice of integer or
odd half-integer spin.  Thus the
requirement that the observable densities commute at equal times, 
except at coincident points,
again leads to the correct association of spin with either c-number canonical 
equal-time commutators or anticommutators.\\

Gauge theories in covariant gauges have a space of states with an indefinite
metric.  Since both the spin-statistics theorem and the spin-locality theorem
assume a positive-definite metric, we have to understand how these theorems can
apply to the particles and fields in gauge theories.  The qualitative answer is
that such gauge theories have a physical space of states (called $H_{phys}$
by Nakanishi and Ojima) that has a positive-definite 
metric. The space $H_{phys}$ is the quotient of a subspace (called 
${\cal V}_{phys}$ 
by Nakanishi and Ojima) of an indefinite metric space (called ${\cal V}$) and
the space of zero norm states (called ${\cal V}_0$ by Nakanishi and Ojima) 
and the theorems presumably hold in this physical space.\\ 

The local observable point of view allows a very general discussion of the
spin-statistics connection based on the principles of locality, relativistic
invariance, and spectrum without reference to fields.  The literature on this
point of view can be traced from the book by R. Haag\cite{local} and the
talk by S. Doplicher\cite{dop}. 

{\flushleft 2. $TCP$}

Now I turn to the $TCP$ theorem\cite{jost}.  
The $TCP$ theorem in Jost's formulation (given for simplicity for a single
charged field) states
that the necessary and sufficient condition for $TCP$ to be a symmetry of the
theory in the sense that there is an antiunitary operator $\theta$ such that 
\beee 
\theta \phi(x) \theta^{-1} = \phi^{\dagger}(x),~~
\theta \phi^{\dagger}(x) \theta^{-1} = \phi(x),~~ \theta |0 \rangle =|0 \rangle,
\eeee
is that the field $\phi$ obey weak local commutativity in a real neighborhood
of a Jost point.  Jost points are the points where all convex sums of 
the successive difference
vectors of the points in a vacuum matrix element are purely spacelike.  Local
commutativity implies weak local commutativity, but weak local commutativity is
much weaker than local commutativity.  I amplify Jost's example\cite{jost2} 
of a free relativistic neutral scalar field quantized with
Fermi statistics, repeat Jost's proof that this field 
obeys the $TCP$ theorem, and 
find the Hamiltonian density for this field.

Expand the field in terms of annihilation and creation operators 
that obey relativistic normalization,
\beee
\phi(x)=\frac{1}{(2 \pi)^3} \int \frac{d^3k}{2E_k}(A(k)exp(-ik \cdot x) 
+A^{\dagger}(k)exp(ik \cdot x)).                              \label{6}
\eeee
The annihilation and creation operators obey
\beee
[A(k),A^{\dagger}(l)]_+=2E_k \delta({\bf k}-{\bf l}),         \label{7}
\eeee
\beee
[A(k),Al)]_+=0,~~[A^{\dagger}(k),A^{\dagger}(l)]_+=0.         \label{8}
\eeee
The anticommutator of the field is
\beee
[\phi(x),\phi(y)]_+=\Delta^{(1)}(x-y),
\eeee
which is not local.  With a vacuum that is annihilated by the annihilation
operators, $A(k)|0\rangle=0$, this is a theory of free, neutral scalar fermions.
This example is nonlocal, but the field does not vanish.  It obeys the $TCP$
theorem, because its vacuum matrix elements are sums of products of two-point
vacuum matrix elements and its two-point vacuum matrix elements obey local
commutativity from the properties of spectrum and Lorentz invariance.  (To
further emphasize how weak a condition $TCP$ invariance is, note that even
free quon fields obey $TCP$.\cite{quon})

The Hamiltonian for this free theory is
\beee
H=\int \frac{d^3k}{2E_k}E_k A^{\dagger}(k)A(k). \label{9}
\eeee
Translate this into position space using
\begin{equation}
A(k)=i\int e^{ik \cdot x} 
\stackrel{\leftrightarrow}{\textstyle \partial}^0\phi
(x) \frac{d^3x}{(2\pi)^{3/2}},                             \label{10}
\end{equation}
\begin{equation}
A^{\dagger}(k)=-i\int e^{-ik \cdot x} 
\stackrel{\leftrightarrow}{\textstyle \partial}^0\phi (x)
\frac{d^3x}{(2\pi)^{3/2}}.                       \label{11}
\end{equation}
The result is
\begin{equation}
H = \int d^3x d^3y \, \, [i \partial^0_x \Delta^{(1)} (x-y)] \stackrel
{\leftrightarrow}{\textstyle \partial}^0_x
\stackrel{\leftrightarrow}{\textstyle \partial}^0_y: \phi^{\dagger} (x)
\phi (y):],                                                \label{14}
\end{equation}
which is the integral of a (nonlocal) energy density,
\beee
H=\int d^3x {\cal H}(x),                                          \label{15}
\eeee
\begin{eqnarray}
{\cal H}(x)&=&i~\int d^3\rho [\stackrel{.}{\Delta}^{(1)}(\rho):
\stackrel{.}{\phi}^{\dagger}(x+\rho/2)\stackrel{.}{\phi}(x-\rho/2): 
-\stackrel{..}{\Delta}^{(1)}(\rho):
\phi^{\dagger}(x+\rho/2)\stackrel{.}{\phi}(x-\rho/2):  + \nonumber \\
& & \stackrel{..}{\Delta}^{(1)}(\rho):
\stackrel{.}{\phi}^{\dagger}(x+\rho/2)\phi(x-\rho/2):
-\stackrel{{\bf ...}}{\Delta}^{(1)}(\rho)
:\phi^{\dagger}(x+\rho/2)\phi(x-\rho/2):].                        \label{16}
\end{eqnarray}
This energy density is nonlocal in both senses: it is not a pointlike functional
of the fields and it does not commute with itself at spacelike separation.  This
result for ${\cal H}(x)$ also follows from Eq.(1).  The difference between the
spin-0 field quantized with Fermi statistics (the wrong case) and with Bose
statistics (the right case) is that for the wrong case the $\Delta^{(1)}(x)$
distribution enters rather than $\Delta(x)$, and the zero-time values of 
$\Delta^{(1)}(x)$ are not local, in contrast to the vanishing of the zero-time
value of $\Delta(x)$ and
the locality of its time derivative at zero time.
\vspace{.25in}

{\it Summary} 

One should distinguish three theorems:
{\it The spin-statistics theorem}:
Given the choice between Bose and Fermi statistics, particles with integer spin
must obey Bose statistics and particles with odd half-integer spin must obey
Fermi statistics.
{\it The spin-locality theorem}:
Given the choice between commutators that vanish at spacelike separation and
anticommutators that vanish at spacelike separation, fields with integer spin
must have local commutators and fields with odd half-integer spin must have
local anticommutators.
{\it The TCP theorem}:
The necessary and sufficient condition for the existence of an antiunitary
operator $\theta$ such that $\theta \phi(x) \theta^{-1}=\phi^{\dagger}(-x)$,
~$\theta \phi^{\dagger}(x) \theta^{-1}=\phi(-x), \theta |0 \rangle =|0 \rangle$,
is weak local commutativity at Jost points.

For the spin-statistics theorem, the basis of the theorem is the requirement
that observables commute at spacelike separation.  If the wrong choice is made,
observable densities fail
to commute at spacelike separation.    For fields that don't have
asymptotic fields, the choice is between fields whose canonical
variables have c-number equal-time commutators and fields whose canonical
variables have c-number equal-time anticommutators.  For the spin-locality
theorem, if the wrong choice is made, the field vanishes.  The $TCP$ theorem 
can hold even if the
field and its particles have the wrong connection of spin and statistics;
clearly it can hold under very general conditions.\\

{\bf Acknowledgements}

I am happy to thank Xiangdong Ji, Jim Swank, and Ching-Hung Woo 
for stimulating questions and helpful suggestions.  I especially thank Joe
Sucher for a careful reading of the text and for many suggestions to improve
this note.  I thank Sergio Doplicher and Kurt Haller for bringing relevant
references to my attention.


\begin{thebibliography}{99}
\bibitem{5} It is also true that these two theorems are
not exclusive theorems; rather the spin-statistics theorem
should include the cases in which Bose is replaced by parabose and Fermi is
replaced by parafermi, in each case of any integer order $p$, with $p=1$ being
the Bose or Fermi case and the spin-locality theorem should include the
corresponding cases of paralocality and para-antilocality\cite{green,dgs}.
\end{thebibliography}
\end{document}